\documentclass[prd, aps, showkeys, twocolumn, superscriptaddress, showpacs, nofootinbib, usenatbib, longbibliography]{revtex4-2}

\usepackage{url}
\usepackage[hidelinks]{hyperref}
\usepackage{tikz}
\usepackage{changes}
\usepackage{subcaption}
\usepackage[normalem]{ulem}
\usepackage[justification=raggedright,singlelinecheck=false]{caption}

\usepackage{graphicx} 
\usepackage{amsmath}
\usepackage{appendix}

\hypersetup{
    colorlinks=true,
    linkcolor=blue!70!black,  
    citecolor=green!50!black, 
    urlcolor=blue!50!white    
}

\newcommand{\Mparamwide}{$\mathcal{M}^{\mathrm{TM:wide}}_{\mathrm{parametric}}$}
\newcommand{\Mparaminfo}{$\mathcal{M}^{\mathrm{TM:info}}_{\mathrm{parametric}}$}
\newcommand{\Magnoswide}{$\mathcal{M}^{\mathrm{TM:wide}}_{\mathrm{agnostic}}$}
\newcommand{\Magnosinfo}{$\mathcal{M}^{\mathrm{TM:info}}_{\mathrm{agnostic}}$}

\date{\today}

\begin{document}
\title{Bayesian P-spline recovery of stochastic gravitational-wave backgrounds in LISA}
\author{Nazeela Aimen}
\affiliation{Department of Statistics, The University of Auckland, Auckland, New Zealand}
\author{Patricio Maturana-Russel}
\affiliation{Department of Statistics, The University of Auckland, Auckland, New Zealand}
\affiliation{Department of Mathematical Sciences, Auckland University of Technology, Auckland, New Zealand}
\author{Avi Vajpeyi}
\affiliation{Department of Statistics, The University of Auckland, Auckland, New Zealand}
\author{Nelson Christensen} 
\affiliation{Universit\'e C\^ote d'Azur, Observatoire de la C\^ote d'Azur, Artemis, CNRS, 06304 Nice, France}
\author{Renate Meyer}
\affiliation{Department of Statistics, The University of Auckland, Auckland, New Zealand}

\begin{abstract}
The detection of a stochastic gravitational-wave background (SGWB) is a primary science objective for the Laser Interferometer Space Antenna (LISA). However, extracting these signals is difficult because both the signal and the instrumental noise are stochastic and overlapping in the millihertz band. In this work, we present a Bayesian framework for the joint estimation of LISA noise and SGWB signals. Our approach models the LISA instrumental noise using flexible log-penalized splines, employing a roughness penalty to prevent overfitting while maintaining computational efficiency. For the SGWB, we compare a power-law model with a spline-based model and study how the choice of signal model and noise prior affects signal recovery and detection. Using simulated LISA data, we find that the power-law model gives tighter estimates when the signal follows the assumed shape. However, it fails to recover a localized spectral feature that is not described by a power law, causing the signal to be absorbed by the instrumental-noise spline. The fully spline-based model is less restrictive and successfully recovers such features. We also find that stronger prior information about the test-mass noise helps reduce the degeneracy between the noise and SGWB models at low frequencies, improving signal detection. These results reflect a single trade-off: added model flexibility reduces sensitivity when the assumed signal shape is correct, and prevents bias when it is not.
\end{abstract}

\maketitle

\section{Introduction}
The first direct detection of gravitational waves (GWs) by ground-based interferometers in 2015 established GW astronomy as a new observational window on the Universe~\cite{Abbott2016_GW150914}. Since then, the search for GWs has expanded across a broad frequency spectrum, utilizing ground-based interferometers, pulsar timing arrays, and future space-based detectors. Beyond individually resolvable sources, a crucial source for current and future detectors is the stochastic gravitational-wave background (SGWB), which is a stochastic signal produced by the incoherent superposition of numerous unresolved GW sources or by processes in the early Universe~\cite{romano_detection_2017,caprini_cosmological_2018, regimbau_astrophysical_2011, van_remortel_stochastic_2023}. In the millihertz band in particular, the SGWB can arise from both astrophysical and cosmological sources. Astrophysical backgrounds arise from the superposition of compact binaries across cosmic time (e.g., stellar-mass binaries or massive black-hole binaries), while cosmological backgrounds may be generated by inflationary mechanisms, first-order phase transitions, or cosmic strings~\cite{caprini_cosmological_2018}. More broadly, SGWBs are being searched for across the whole GW spectrum. Ground-based laser interferometers~\cite{abbott_directional_2017,abbott_upper_2017,abbott_gw170817_2018,abbott_search_2019,abbott_directional_2019,abbott2021upper,abbott2021search,abbott2022all,abac2025cosmological,abac2025upper} and pulsar timing arrays~\cite{arzoumanian_nanograv_2020,arzoumanian_searching_2021,antoniadis_international_2022,agazie2023nanograv,antoniadis2023second} are actively searching for such backgrounds in their respective frequency bands. Moreover, future ground-based~\cite{abbott2017exploring,maggiore2020science,branchesi2023science} and space-based detectors~\cite{kawamura2011japanese,baker2019space,kawamura2021current,komori2025current} will join this quest. This is one of the major objectives for the Laser Interferometer Space Antenna (LISA)~\cite{amaro2017laser}.

As a future space-based GW observatory, LISA is designed to access the millihertz GW band, enabling observations of sources that are inaccessible to ground-based detectors~\cite{amaro2017laser,amaro2023astrophysics}. LISA will consist of three spacecraft in a triangular heliocentric constellation with arm lengths of the order of millions of kilometers. In this frequency range, LISA is expected to observe deterministic signals from compact galactic binaries, massive black-hole mergers, extreme mass-ratio inspirals, and stochastic backgrounds from both astrophysical and early-Universe sources~\cite{amaro2017laser,amaro2023astrophysics, caprini_cosmological_2018}.

Extracting a SGWB with LISA is difficult because both the SGWB and the instrument noise are stochastic in nature, and LISA cannot cross-correlate two independent detectors the way ground-based interferometers can~\cite{romano_detection_2017,christensen_stochastic_2019}. This challenge motivates the present work, in which we focus on separating the SGWB from LISA instrument noise. Several studies have used a simple structure for the SGWB, often a power law, and then inferred a small number of parameters, such as an amplitude and slope~\cite{boileau_spectral_2021}. As the LISA band contains a large number of overlapping sources (especially galactic binaries), several pipelines aim for a ``global fit'', where numerous signals and the noise and foreground are fitted together~\cite{lackeos_lisa_2023,strub_global_2024}. Many previous analyses assumed parametric models for both the LISA noise and the SGWB signal~\cite{adams_detecting_2014,robson_construction_2019,boileauProspectsLISADetect2023,liang2026fullcovariancebayesianinferencestochastic}. This would work in the ideal scenario where the noise assumptions are met perfectly. However, we expect the real noise spectrum to be more complex and therefore have high chances of differing from the assumed shape, which could bias SGWB results \cite{muratoreImpactNoiseKnowledge2023}. 

To reduce this dependence on a fixed noise shape, more flexible (non-parametric or weakly parametric) noise models have been proposed in the literature. 
One such example is the spline-based approach of \cite{baghi_uncovering_2023}, where the LISA noise spectrum is modeled non-parametrically across frequency, while the SGWB signal is assumed to follow a power law. Along the same lines, \citep{karnesis_effect_2026} compares parametric and flexible models for both the instrumental noise and the SGWB signal, and maps the resulting detectability bounds in the amplitude--spectral-index plane using Bayes factors between noise-only and noise-plus-signal hypotheses.
Their flexible spectra are built from Akima splines whose number and knot positions are inferred with trans-dimensional reversible jump MCMC, and they model the noise at the individual interferometric links before propagating it through the TDI transformations, retaining the full $3\times3$ spectral density matrix of the $XYZ$ channels. Similarly, \cite{pozzoli_weakly_2024} uses smooth stochastic process models based on Gaussian processes to adapt to unknown spectral features while enabling joint inference of the SGWB. All these models assume perfect knowledge of the LISA noise transfer functions and the SGWB response. In contrast, \cite{santini_flexible_2025}
introduces a two-step inference strategy where a parametric estimation of both noise and SGWB is followed by a second analysis where any remaining spectral structure not captured by the parametric model is modeled by a linear combination of splines. Separately, recent work extends Bayesian P-spline spectral estimation to the multivariate setting, enabling flexible modeling of the full TDI spectral density matrix, including cross-channel correlations~\cite{2026arXiv260704833V}. These studies introduce flexibility in different parts of the analysis, either in the instrumental-noise model, the SGWB model, or through an additional inference step. In this work, to address potential biases in signal detections introduced by parametric signal or noise assumptions, we propose a flexible modeling framework for both the LISA instrumental noise and the SGWB signal. 

Building on earlier P-spline work~\cite{aimenBayesianPowerSpectral2026,maturana-russel_bayesian_2021}, we model LISA noise using a large number of B-spline basis functions across the frequency domain, with a penalty prior imposed on the spline weights to prevent overfitting. In this work, we extend our previous noise-only estimation to a joint framework for noise and signal parameter estimation, enabling signal separation within the LISA $A$, $E$, and $T$ channels. Specifically, we assign log-Penalized-spline (log-P-spline) priors on the two primary instrumental noise sources—test-mass and readout—and propagate them through the TDI transfer functions to obtain the channel noise power spectral densities (PSDs). For the SGWB, we compare two signal models: a standard parametric power law and a flexible log-P-spline model. We combine each signal model with two data-informed priors of different levels of informativeness on the test-mass noise spline, which allows us to assess the impact of model flexibility on SGWB recovery and detection sensitivity in future LISA data.

Our approach differs from~\cite{karnesis_effect_2026} in three respects. First, we fix the number of splines in advance and control overfitting and smoothness using a roughness-penalty prior, thereby avoiding trans-dimensional sampling. Second, we work with the orthogonal $AET$ channels, for which the spectral-density matrix is diagonal under our assumptions. The $T$ channel is treated as an approximate null channel for the SGWB and therefore provides additional information on the instrumental noise, particularly the OMS component, which helps separate the SGWB from the noise. Third, their SGWB injections are power laws. In contrast, we also inject an SGWB signal with a spectral shape not included in the parametric model used for fitting, a regime in which a flexible signal model is expected to matter most.

Our methodology has several important advantages for LISA SGWB analysis. By modeling the instrumental test-mass and readout noise, and SGWB signal spectra with log-P-splines, the method can adapt to deviations from parametric shapes, reducing the risk of bias from model misspecification. To test this flexibility, we also analyze a Gaussian-bump injection whose shape is not captured by the power-law model. Moreover, the use of roughness-penalty priors allows a large spline basis while controlling overfitting, thereby keeping the model flexible without compromising smoothness. Because the dimension of the spline model is fixed at the beginning, the method avoids the need for reversible-jump Markov chain Monte Carlo to explore models with varying numbers of basis functions or knots. This computational efficiency is especially valuable for long LISA data sets, where fast, efficient frequency-domain inference is required. 
        
The paper is structured as follows: in Section~\ref{mod}, we describe the models used in this work, including the SGWB power-law model, the LISA test-mass and readout-noise models, and the TDI transfer functions. In Section~\ref{sec:methodology}, we describe our framework, including the blocked Whittle likelihood for the averaged periodograms in the $A$, $E$, and $T$ channels, the log-P-spline priors for the test-mass and OMS noise PSDs, and the priors for the SGWB parameters. Section~\ref{Results} presents our results on simulated data, evaluating pipeline performance on both a standard power-law background and an unmodeled Gaussian bump feature. Finally, in Section~\ref{conclusion}, we summarize the main findings and discuss future work.

\section{Models}\label{mod}
\subsection{Stochastic gravitational wave background}
A SGWB is produced by the superposition of a large number of unresolved sources and may contain both astrophysical and cosmological contributions. To evaluate how our framework handles different physical scenarios, we consider two distinct spectral shapes of the signal in this work. First, we assume the energy density per logarithmic frequency intervals of an isotropic SGWB $\Omega_{\rm gw}(f)$, parametrized as a power law around a pivot frequency $f_{0}=3.16\,\text{mHz}$~\cite{baghi_uncovering_2023}
\begin{equation}
\Omega_{\rm gw}(f) = \Omega\left(\frac{f}{f_{0}}\right)^{\alpha},
\label{eq:omega_powerlaw}
\end{equation}
where $\Omega$ is the energy density at the pivot frequency $f_{0}$ and $\alpha$ is the spectral index. Second, to test the robustness of our framework against signals that deviate from a smooth power law, such as those expected from primordial black hole formation or first-order phase transitions, we also analyze a localized, log-normal Gaussian bump feature~\cite{capriniReconstructingSpectralShape2019}. The explicit mathematical formulation and parameters for this localized feature are detailed in Section~\ref{gaussian_bump}.

$\Omega_{\rm gw}(f)$ is related to the (one-sided)
power spectral density $S_h(f)$ by~\cite{caprini_cosmological_2018}
\begin{equation}
\Omega_{\rm gw}(f) = \frac{4\pi^2}{3H_0^2}\, f^3\, S_h(f),
\label{eq:omega_sh}
\end{equation}
where $H_0$ is the Hubble constant. 

The spectral density of the SGWB in the TDI channel $C$ is given by
\begin{equation}
S_{C,\rm gw}(f) = R_C(f)\, S_h(f), \quad C\in \{A,E,T\}
\label{eq:lisa_sgw}
\end{equation}
where $R_C(f)$ is the sky- and polarization-averaged LISA response function for channel $C$. In this analysis, we assume the response is perfectly known and follow the formulation provided in~\cite{baghi_uncovering_2023} to calculate it. We generate the SGWB strain in the time domain using the inverse Fourier-transform method.
\subsection{LISA noise}\label{subsec:lisanoise}
The sensitivity of LISA is limited by two primary instrumental noise sources: test-mass (TM) acceleration noise and optical metrology system (OMS) readout noise. Following the instrument specifications in \cite{bayle_unified_2023}, the acceleration noise power spectral density is modeled as
\begin{equation}\label{eq:stm} 
        S_{\rm TM}(f)=a^2_{\rm TM}\Big[ 1+\Big(\frac{f_1}{f}\Big)^2 \Big]\Big[1+\Big(\frac{f}{f_2}\Big)^4\Big]
        \Big(\frac{1}{2\pi fc}\Big)^2,
\end{equation}
where 
$a_{\rm TM}=3\times 10^{-15}\mathrm{ms}^{-2}$, $f_1=4\times 10^{-4}\,\mathrm{\mathrm{Hz}}$, $f_2=8\times10^{-3}\,\mathrm{\mathrm{Hz}}$, 
and the readout noise, which dominates at higher frequencies, is given by
\begin{equation}
    \label{eq:oms}
    S_{\rm OMS}(f)=a^2_{\rm OMS}\Big[ 1+\Big(\frac{f_3}{f}\Big)^4 \Big]\Big(\frac{2\pi f }{c}\Big)^2,
\end{equation}
where $a_{\rm OMS}=15\times 10^{-12}\mathrm{ms}^{-2}$, $f_3=2\times 10^{-3}\,\mathrm{Hz}$. 
\subsubsection{TDI}\label{subsec:TDI}
To suppress laser frequency noise, we employ second-generation Time Delay Interferometry (TDI). We adopt the orthogonal $A, E$, and $T$ channels defined in~\cite{quang_nam_time-delay_2023}. 

The $A$ and $E$ channels are sensitive to GW signals, while the $T$ channel, often referred to as the `null' channel, is primarily noise dominated, which makes it significant for noise characterization~\cite{princeLISAOptimalSensitivity2002}. Assuming equal arm lengths $L_{\rm arm}=2.5\,\text{Gm}$ and uncorrelated noise across links, the noise PSDs for these channels are constructed from the underlying TM and OMS contributions.

The total noise in the $A$ (and by symmetry $E$) channel is
\begin{align}
    S_{A,\rm noise}(f)=&2C_{X}(\omega)\Big([3+2\cos\omega +\cos2\omega ]2S_{\rm TM}(f)\\
    \nonumber
    &+[2+\cos\omega]S_{\rm OMS}(f) \Big),
\end{align}
while the $T$ channel noise is
\begin{align}
    S_{T,\rm noise}(f)=&4C_{X}(\omega)\Big(8\sin^4\Big(\frac{\omega}{2}\Big)S_{\rm TM}(f)\\
    \nonumber
    &+[1-\cos\omega]S_{\rm OMS}(f)\Big),
\end{align}
where,
\begin{align}
    \omega &= 2\pi f\frac{L_{\rm arm}}{c},\\
    C_{X}(\omega) &= 16 \sin^2(\omega )\,\sin^2(2\omega).
\end{align}
This formulation assumes perfect laser-noise cancellation and ignores correlations arising from arm length mismatches. While equations \eqref{eq:stm} and \eqref{eq:oms} provide the theoretical instrument noise models, in Section~\ref{subsec:noiseprior} we introduce a P-spline framework to account for potential deviations from these analytic forms.
\section{Methodology}\label{sec:methodology}
Consider a stationary time series $\mathbf{Z}$ of length $n_s$, partitioned into $J$ mean-centered segments, $Z^j = (Z_0^j,\ldots,Z_{N-1}^j)$, each of a duration $T=N\Delta_t$. We have sampling interval $\Delta_t$, which determines the sampling frequency $f_s=1/\Delta_t$ and the Nyquist limit $f_{\mathrm{Ny}}=1/(2\Delta_t)$.

For each segment, we apply a window \(w=(w_0,\ldots,w_{N-1})\), with $U=\sum_{t=0}^{N-1} w_t^2 .$ The resulting windowed-discrete Fourier transform for the $j$-th segment is then expressed as
\begin{equation}
X_j(f_l)=\sum_{t=0}^{N-1}w_t Z_t^j\exp\left(-i2\pi f_l t\Delta_t\right),    
\end{equation}
where the Fourier frequencies are $f_l=l/(N\Delta_t)$ for $l = 0, \ldots, v,$ with $\,v= N/2-1$  when $N$ is even and $v=(N-1)/2$ when $N$ is odd. The corresponding window-normalized Fourier coefficient is $d_j^{(w)}(f_l)
=X_j(f_l)/\sqrt{f_s U}$, which allows for the construction of the windowed periodogram
\begin{equation}
I_j^{(w)}(f_l)
= d_j^{(w)}(f_l)d_j^{(w)}(f_l)^*
=\frac{|X_j(f_l)|^2}{f_sU},
\end{equation}
For a sufficiently large number of points $N$, the covariance of these window-normalized coefficients is approximately equal to the two-sided continuous spectral density ($\text{Cov}\left(d_j^{(w)}(f_l)\right)\approx S_j(f_l)$), provided the underlying spectrum remains smooth across the bandwidth of the window~\cite{dahlhausSmallSampleEffects1988}. Assuming that these $J$ segments are independent, and that they share an identical power spectral density $S(f_l)$, the sum of these periodograms follows approximately a Gamma distribution with shape parameter $J$. Hence, the blocked Whittle likelihood for three TDI channels is~\cite{whittleCurvePeriodogramSmoothing1957}
\begin{align}
\log L(\mathbf{Z}\mid S)=-J\sum_{C}\sum_{l=1}^{v}\left[\log S_C(f_l)+\frac{\bar I_C(f_l)}{S_C(f_l)}\right],
\end{align}
where, 
\begin{align}\label{avgpdgrm}
\bar{I}_C(f_l)=\frac{1}{J}\sum_{j=1}^J I_{C,j}(f_l), \quad \quad C\in \{A,E,T\}
\end{align}
denotes the averaged periodogram. Windowing can introduce additional correlations between neighboring Fourier frequencies, so the frequency-bin independence assumed by the Whittle likelihood is approximate.

The total power spectral density in each channel is a combination of the instrumental noise $S_{C,\rm noise}$ and the SGWB signal $S_{C,\rm gw}$. 
As described in Section~\ref{subsec:TDI}, we assume the $T$ channel serves as a null channel with negligible signal sensitivity, while the $A$ and $E$ channels contain the primary signal contributions. Thus,
\begin{align}
    &S_A(f)=S_{A,\rm noise}(f)+S_{A,\rm gw}(f),\\
    &S_E(f)=S_{E,\rm noise}(f)+S_{E,\rm gw}(f),\\
    &S_T(f)=S_{T,\rm noise}(f).
\end{align}
To handle unexpected differences from theoretical noise models and to allow for a flexible signal reconstruction, we use a Bayesian framework. In the following subsections, we define the log-P-spline priors used to model the components of both the instrumental noise and the SGWB signal.
\subsection{Noise priors}\label{subsec:noiseprior}
To accommodate deviations from the theoretical instrument models, we place log-P-spline priors on both the test-mass acceleration and OMS noise components. Specifically, we model the logarithm of each noise PSD as a linear combination of $K$ B-spline basis functions
\begin{equation}
\log(S_i(f))=\sum_{k=1}^{K}\lambda_{k,i} b_{k,r,i}(f;\boldsymbol{\xi}_{i}), \quad i \in {\text{TM, OMS}},    
\end{equation}
where, for each component $i \in \{\text{TM, OMS}\}$, $b_{k,r,i}$ is a B-spline basis function of fixed degree ($r$), normalized to have unit integral over frequency. In this work, we use cubic B-splines, corresponding to $r=3$. Following the convention in our previous work~\cite{aimenBayesianPowerSpectral2026}, the knot sequence $\boldsymbol{\xi}_i=\{f_{min}=\xi_{0,i}=\xi_{1,i}=\xi_{r,i},\leq \xi_{r+1,i}\leq \dots \leq \xi_{K,i}=\xi_{K+1,i}=\dots=\xi_{K+r,i}=f_{max}\}$ is fixed over the frequency range $[f_{\min}, f_{\max}]$ of the periodogram. The number of basis functions $K$ is calculated by $K = N_\xi + r - 1$, where $N_\xi$ is the number of knots (which includes the boundary knots). To effectively capture the variations in the LISA noise, we utilize B-spline bases defined on logarithmically spaced knots for both noise components.

The spline coefficients are denoted by the vector $\boldsymbol{\lambda}_i = (\lambda_{1,i}, \dots, \lambda_{K,i})^\top$. To incorporate prior information regarding the expected LISA instrument performance, we center the prior distributions for these weights on the theoretical noise spectra defined in Eq.~\eqref{eq:stm} and Eq.~\eqref{eq:oms}. These models are used to obtain the central weight vectors, $\boldsymbol{\lambda}_{\text{loc,TM}}$ and $\boldsymbol{\lambda}_{\text{loc,OMS}}$, which define the prior mean. Specifically, we obtain the central weight vectors by fitting the B-spline basis to the logarithm of the corresponding theoretical parametric spectra using penalized least squares. We then specify a multivariate normal prior for the weights:
\begin{equation}\label{eq:lam_pri_noise}
\boldsymbol{\lambda}_i \mid \phi_i \sim \mathcal{N}_{K}\left(\boldsymbol{\lambda}_{\text{loc},i}, (\phi_i \mathbf{P}_i)^{-1}\right), \quad i \in {\text{TM, OMS}},
\end{equation}
where $\mathbf{P}_i$ is a full-rank penalty matrix that ensures spectral smoothness; for a detailed construction of this matrix see~\cite{aimenBayesianPowerSpectral2026}. 
The precision parameter $\phi$ regulates the smoothness of the spline coefficients: higher values promote a smoother spectral reconstruction, whereas smaller values permit greater local flexibility.

The choice of the precision parameter $\phi$ is crucial for balancing model flexibility through the smoothness of the spline coefficients. For the OMS noise component, we adopt a wide prior on the weights by setting $\phi_{\rm OMS} = 10^4$. However, for the test-mass acceleration noise, we evaluate two distinct prior configurations: a wide prior ($\phi_{\rm TM} = 10^4$) and an informed prior ($\phi_{\rm TM} = 10^8$). We test them to assess the efficiency of signal detection across both parametric and agnostic SGWB signal models. Appendix~\ref{appendix: phi} illustrates how the choice of $\phi$ affects the spread of the PSDs drawn from the prior.

A primary motivation for investigating an informed test-mass prior is the identifiability issue at low frequencies. As demonstrated in Appendix~\ref{appendix: sensitivity}, the TDI $T$ channel is weakly sensitive to test-mass acceleration noise, meaning the $T$ data provides negligible information to distinguish $S_{\rm TM}(f)$ from a low-frequency-dominant stochastic signal. This degeneracy is particularly problematic when the SGWB is modeled using log-P-splines, as the lack of parametric assumptions about the signal allows the noise spline to absorb all variations. By applying a tight constraint on the test-mass spline ($\phi_{\rm TM} = 10^8$), we restrict the noise to its theoretical expectation, thus increasing the probability of signal detection.
\subsection{Signal priors}
For the SGWB signal, we test two different priors. The first is a parametric power-law template, which matches the model used for our data simulations:
\begin{equation}
    S_h(f) = \frac{3H_0^2}{4\pi^2f^3} \Omega\left(\frac{f}{f_{0}}\right)^{\alpha},
\end{equation}
with wide priors $\log_{10}\Omega\sim \mathcal{U}[-20,-9]$ and $\alpha\sim\mathcal{U}[-5,5]$. These distributions ensure that the analysis remains sensitive to all theoretically detectable signals within the considered parameter space. 

The second prior we assign is a log-P-spline prior defined on the log-spectral density of the signal. This allows for greater flexibility in capturing spectral shapes that deviate from a pure power law. The model is expressed as
\begin{equation}
\log(S_h(f))=\sum_{k=1}^{K}\lambda_{k,\text{gw}} b_{k,r,\text{gw}}(f;\boldsymbol{\xi}_{\text{gw}}).
\label{gwpsp}
\end{equation}

Following the methodology described in the previous section, we establish the basis functions $b_{k,r,\text{gw}}$ and the knot sequence $\boldsymbol{\xi}_{\text{gw}}$ for the signal. To ensure that the prior is based on physical phenomena in the universe, we center the weights on a reference power-law model with $\alpha=1$ and $\Omega=10^{-13}$. As noted in~\cite{capriniReconstructingSpectralShape2019}, indices in the range of $0.5 \leq \alpha \leq 1$ are of particular interest as they may characterize specific early-universe phenomena, such as a kinetic energy-dominated phase. This reference defines the center of the prior for spline coefficients, $\boldsymbol{\lambda}_{\text{loc,gw}}$. The prior distribution for the weights is then formulated as a multivariate normal distribution:
\begin{equation}\label{eq:lam_pri}
    \boldsymbol{\lambda}_{\rm gw} \mid \phi_{\rm gw} \sim \mathcal{N}_{K}(\mathbf{\boldsymbol{\lambda}_{\rm loc, gw}},(\phi_{\rm gw} \mathbf{P_{\rm gw}})^{-1}),
\end{equation}
where $\mathbf{P}_{\rm gw}$ is the penalty matrix for the SGWB signal. We adopt a very wide prior by setting the precision parameter to $\phi_{\rm gw} = 10$, providing a weak constraint that allows the spline to freely explore and capture the signal spectral features.
\subsection{Model selection}
To determine whether the data support the presence of an SGWB, we compare two competing Bayesian models. The first assumes purely instrumental noise, while the second includes both instrumental noise and an SGWB component: 
\begin{align}
\mathcal{H}_0 &: \text{instrumental noise only},\\
\mathcal{H}_1 &: \text{instrumental noise + SGWB}.
\end{align}
We quantify the relative support for these hypotheses using the Bayes factor, which is based on their marginal likelihoods:
\begin{equation}\label{eq:bayesfact}
\log_{10} B = \log_{10} \mathcal{Z}_1 - \log_{10} \mathcal{Z}_0,
\end{equation}
where $\mathcal{Z}_i$ is the marginal likelihood (or evidence) for model $\mathcal{H}_i$. 
A detection claim for an SGWB signal is made when the Bayes factor exceeds a predefined threshold. In practice, this threshold is chosen either based on conventional interpretations of Bayes factors from previous studies~\cite{kassBayesFactors1995} or calibrated empirically using repeated simulated data realizations, as done in previous SGWB analyses~\cite {adamsDiscriminatingStochasticGravitational2010,karnesisAssessingDetectabilityStochastic2020,baghi_uncovering_2023,pozzoliStochasticSignalReally2025}.
Estimating the Bayes factor is computationally challenging because it requires evaluating the marginal likelihood, a high-dimensional integral over the model parameter space. Since this integral generally has no analytical solution, numerical methods are required. 
This challenge has motivated evidence-estimation techniques such as thermodynamic integration \cite{lartillotComputingBayesFactors2006}, stepping-stone sampling \cite{xieImprovingMarginalLikelihood2011,maturana-russelSteppingstoneSamplingAlgorithm2019}, and related post-processing approaches~\cite{zahraouiGeneralizedSteppingstoneSampling2024,zahraoui_morphz_2025}.\newline
In this work, we estimate the marginal likelihood of each model using stepping-stone sampling~\cite{xieImprovingMarginalLikelihood2011}. The method constructs a sequence of tempered distributions that gradually transition from the prior to the posterior and expresses the evidence as a product of ratios of normalizing constants between adjacent distributions. These ratios are estimated using samples obtained across a ladder of inverse temperatures, providing estimates of $\mathcal{Z}_0$ and $\mathcal{Z}_1$. The Bayes factor in Eq.~\eqref{eq:bayesfact} is then computed from these estimates. The uncertainty of the marginal likelihood estimates is quantified using the moving block bootstrap described in \cite{maturana-russelSteppingstoneSamplingAlgorithm2019} and subsequently propagated to obtain uncertainty estimates for the Bayes factors.

\section{Simulations and results}\label{Results}

We consider an equilateral LISA constellation with equal arm lengths and assume that the data are stationary. We further assume that the test-mass and OMS noise spectra are time-independent and identical across the spacecraft, which simplifies the computation. To quantify the strength of each injected SGWB signal, we compute the
optimal signal-to-noise ratio (SNR) following Ref.~\cite{smithLISACosmologistsCalculating2019}
\begin{equation}
\mathrm{SNR} = \left[T_{\rm obs}\sum_{C=A,E} \int_{0}^{\infty}\text{d}f\frac{S_{C,\rm gw}(f)^2}{S_{C,\rm n}(f)^2}\right]^{1/2},
\label{eq:snr}
\end{equation}
where $T_{\rm obs}$ is the observation time.

We inject 180 power-law SGWB signals on a $12\times15$ grid, using 12 logarithmically spaced values of $\Omega$ between $10^{-15}$ and $5\times10^{-12}$ and 15 linearly spaced values of $\alpha$ between $-3$ and $3$, covering a range relevant to the LISA sensitivity band~\cite{bartoloProbingAnisotropiesStochastic2022}.

For each injected parameter combination, we analyze the same signal-plus-noise data under two competing hypotheses: an instrumental-noise-only model and a model containing both instrumental noise and an SGWB. The corresponding evidence is used to calculate Bayes factors and assess signal detectability under the different modeling assumptions. 

We model the instrumental test-mass and OMS noise components using the log-P-spline priors described in Section~\ref{subsec:noiseprior}. For the SGWB, we consider either a parametric power-law model or a flexible log-P-spline model. We combine these two SGWB models with either a wide or an informed prior on the test-mass noise spline, resulting in four model configurations:
\begin{enumerate}
    \item \textbf{Parametric-wide (\Mparamwide):} 
    A parametric power-law SGWB model with a wide prior on the hyperparameter of the test-mass noise spline coefficients.
    \item \textbf{Parametric-informed (\Mparaminfo):} 
    A parametric power-law SGWB model with an informed prior on the hyperparameter of the test-mass noise spline coefficients.
    \item \textbf{Agnostic-wide (\Magnoswide):} 
    A log-P-spline SGWB model with a wide prior on the hyperparameter of the test-mass noise spline coefficients.
    \item \textbf{Agnostic-informed (\Magnosinfo):} 
    A log-P-spline SGWB model with an informed prior on the hyperparameter of the test-mass noise spline coefficients.
\end{enumerate}
In all four configurations, we use the same wide prior for the OMS noise spline. We use $10$ logarithmically spaced frequency knots for each of the TM, OMS, and SGWB splines, resulting in $K=12$ spline coefficients per component. All three components are modeled using cubic B-splines with a first-order difference penalty.

We perform Bayesian inference using the No-U-Turn Sampler (NUTS)~\cite{HoffmanGelman2014}, an adaptive Hamiltonian Monte Carlo algorithm, as implemented in \texttt{NumPyro}~\cite{phanComposableEffectsFlexible2019}. \texttt{NumPyro} uses \texttt{JAX}~\cite{jax2018github} for automatic differentiation and just-in-time compilation. We first examine spectral recovery for a power-law injection, then compare detection boundaries using Bayes factors, and finally test the models on a non-power-law Gaussian-bump injection.
\subsection{Spectral density estimates} \label{results: spectral density}
\begin{figure*}[ht]
    \centering
    \includegraphics[width=\textwidth]{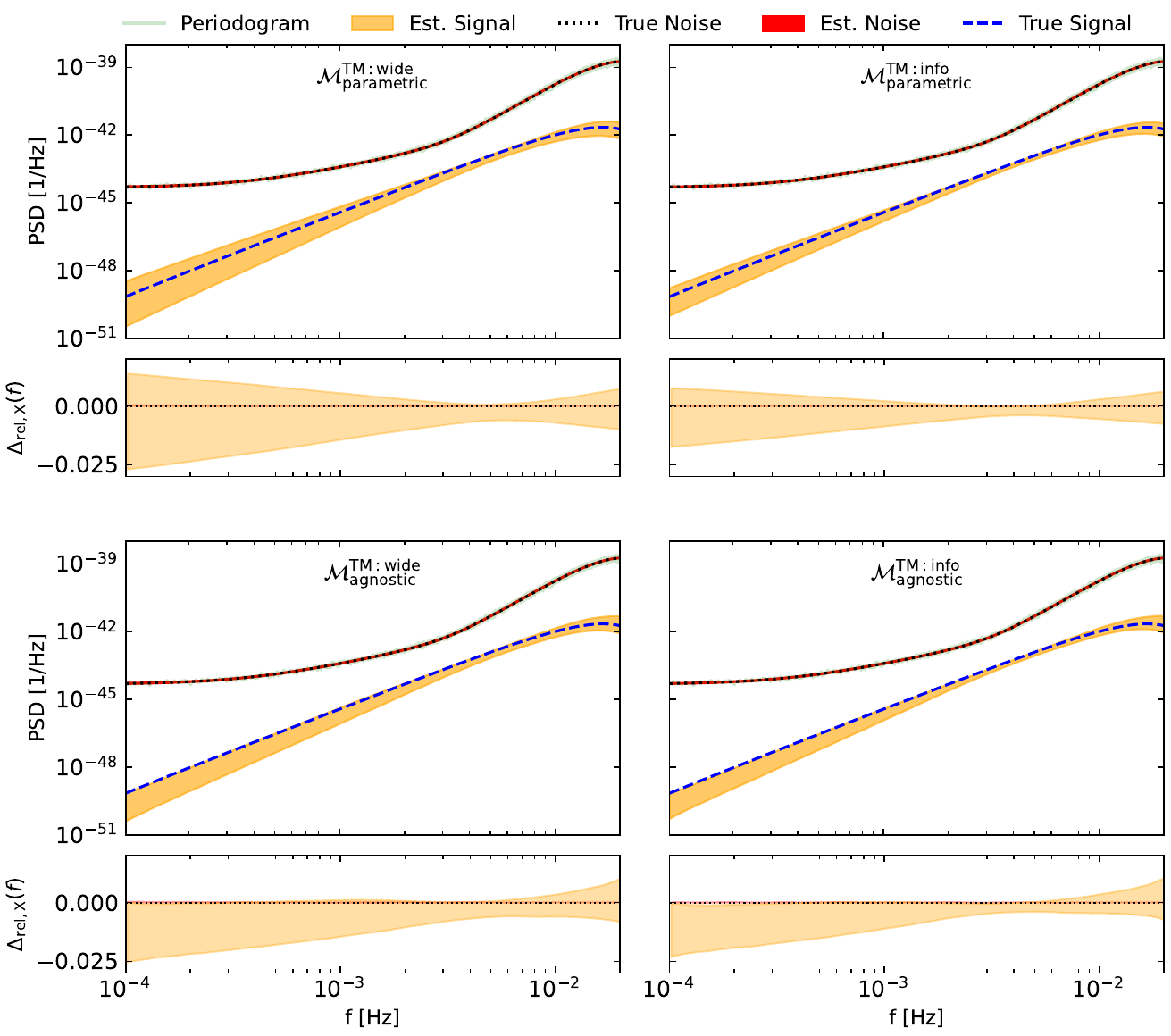}
    \caption{Posterior power spectral density (PSD) estimates and relative errors for the four model configurations, evaluated using a power-law SGWB injection ($\Omega=7\times10^{-14}$, $\alpha=2/3$). Main panels display the averaged $A$-channel periodogram (gray), the $90\%$ posterior credible intervals for the estimated noise (red) and signal (orange), and their true spectra (dashed black and blue). Lower panels show the corresponding relative log-errors for the $5\%$ and $95\%$ posterior quantiles, centered around zero error (dotted line).
    }
    \label{fig:spectraldensityanalysis}
\end{figure*}
For each TDI channel, we partition the one-year LISA time series, sampled at $\Delta_t=2~\mathrm{s}$, into $J=73$ non-overlapping segments of five days each. We apply a Tukey window with shape parameter $\eta_{\rm Tukey}=0.25$ to each segment. This light taper limits spectral leakage while introducing only weak correlations between neighboring Fourier bins. We focus the analysis on the frequency range $f_{\min}=10^{-4}~\mathrm{Hz}$ to $f_{\max}=2\times10^{-2}\,\mathrm{Hz}$, which covers the part of the LISA band with the greatest sensitivity to an SGWB~\cite{robson_construction_2019}. We run four independent Markov chains, each with $3000$ iterations, the first 2000 iterations are used for warm-up and discarded. The remaining $4000$ samples are used for the final inference.

For this analysis, we inject a power-law SGWB of SNR $18.3$ with amplitude $\Omega=7\times10^{-14}$, and spectral index $\alpha=2/3$. This spectral shape is commonly used for the astrophysical background produced by unresolved compact binaries, for which the energy density follows $\Omega_{\rm gw}(f)\propto f^{2/3}$ in the inspiral regime \cite{zhuGravitationalWaveBackground2013,abbott2021upper}. While such compact-binary backgrounds are an important target for ground-based SGWB searches with detectors such as LIGO, Virgo, and KAGRA \cite{abbott2021upper,abac2025upper}, they are also relevant in the mHz band probed by LISA, where unresolved stellar-origin compact binaries can contribute to an astrophysical stochastic background \cite{baghi_uncovering_2023}. Although several astrophysical population models predict larger amplitudes for this type of background~\cite{babakStochasticGravitationalWave2023,lehoucqAstrophysicalUncertaintiesGravitationalwave2023}, we deliberately adopt a relatively weak injection to test the performance of our pipelines under challenging detection conditions. 

Figure~\ref{fig:spectraldensityanalysis} displays the posterior power spectral density (PSD) estimates obtained from the four pipeline variants. The dashed black and blue curves display the theoretical $A$-channel noise and the injected SGWB signal spectra, respectively, while the gray trace shows the averaged periodogram of the $A$-channel. The first row presents the hybrid parametric configurations, where the instrumental noise is modeled using P-splines, while the SGWB signal is described by a parametric power-law model. The second row shows the fully agnostic formulation, where we assign a log-P-spline prior on both the instrumental noise and the SGWB signal. The left column uses the wide TM-noise prior, while the right column uses the informed TM-noise prior.

Across all four cases, the instrumental noise is recovered accurately. This is expected because the $A$-channel periodogram is dominated by the instrumental noise, so the likelihood strongly constrains the noise spectrum. In contrast, the injected SGWB lies well below the instrumental noise over most of the frequency range. Therefore, the signal is only strongly informed by the data in the frequency region approximately between $10^{-3}$ and $10^{-2}\,\mathrm{Hz}$.

The hybrid parametric models recover the injected SGWB more accurately across the full frequency range. This is because the power-law model imposes a fixed spectral shape, allowing information from the sensitive frequency region to constrain the signal amplitude and slope over the entire band. By contrast, the fully agnostic SGWB model has more flexibility and therefore less ability to extrapolate into frequency regions where the signal is weakly constrained by the data. As a result, the spline-based signal estimate follows the injected spectrum mainly in the sensitive part of the band, while it drifts away from the truth at low frequencies where the posterior is more strongly influenced by the prior and the smoothness penalty.

The lower panel of each subfigure shows the relative error,
\begin{equation}
\Delta_{\text{rel},\text{X}}(f) =
\frac{q_{\pm,\text{X}}(f)-\log S_{\text{true},\text{X}}(f)}
{|\log S_{\text{true},\text{X}}(f)|},
\end{equation}
where $q_{\pm,\text{X}}(f)$ denotes the $5\%$ and $95\%$ posterior quantiles of the log-PSD, and $S_{\text{true},\text{X}}(f)$ is either the analytical $A$-channel noise spectrum or the injected SGWB signal spectrum with $\text{X}\in\{\text{n},\text{SGWB}\}$. Values close to zero indicate a closer estimate to the true spectrum.

The informed prior on the test-mass noise spline narrows the posterior uncertainty and helps reduce the degeneracy between the instrumental noise and the SGWB in the sensitive part of the band. However, the SGWB remains weakly constrained outside this region, particularly for the agnostic signal model. The advantages of the formulation when the true signal does not follow a simple power law are examined further in Section~\ref{gaussian_bump}.

For the injection shown in Fig.~\ref{fig:spectraldensityanalysis}, the wide-prior models required approximately $2$ minutes of wall-clock time, while the informed-prior models took less than $20$ minutes. These runs were performed on the OzSTAR supercomputing facility at Swinburne University of Technology using four CPU cores. The longer runtime for the informed prior is likely associated with the tighter constraint on the test-mass spline coefficients, which concentrates the posterior and can make the NUTS sampling more computationally demanding. Runtimes vary across injections and depend on the hardware and model configuration. All four models showed satisfactory sampler convergence, with $\hat{R}$ values close to one and the effective sample sizes sufficiently large. 
\subsection{Bayes factors}\label{results: bayes factors}
\begin{figure}
    \centering
    \includegraphics[width=\columnwidth]{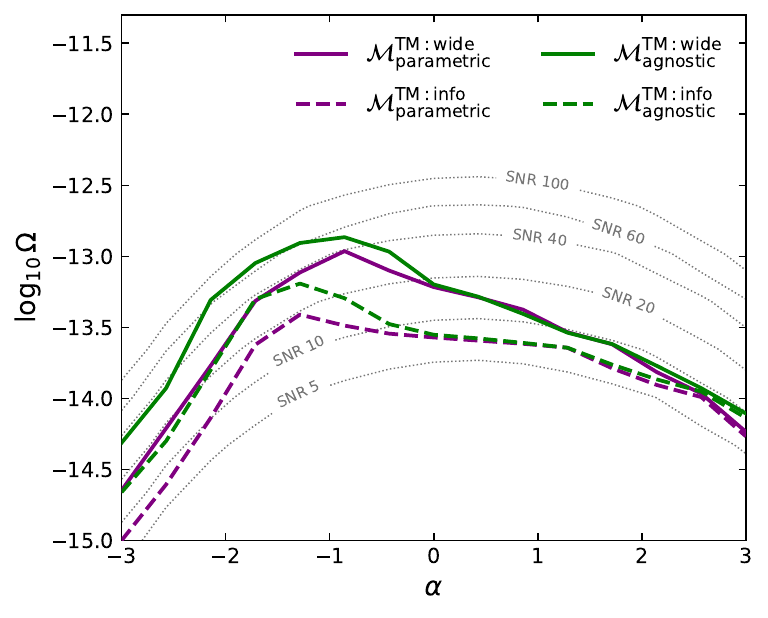}
    \caption{Detection boundaries for the four SGWB analysis configurations as a function of spectral index $\alpha$, defined by $B_{\mathrm{thresh}}=100$ ($\log_{10}B=2$). Solid (dashed) curves correspond to the wide (informed) TM-noise prior, while purple (green) curves correspond to the parametric power-law (agnostic log-$P$-spline) SGWB model. Dotted gray contours indicate constant SNR.
}
    \label{fig:bayes}
\end{figure}
For the Bayes-factor calculation, we use the expectation value of the data rather than individual realizations, following~\cite{baghi_uncovering_2023,karnesis_effect_2026}. This provides a conservative criterion for detection for a fixed detection threshold $B_{\mathrm{thresh}}$. We adopt $B_{\mathrm{thresh}}=100$, corresponding to $\log_{10}B=2$, as the threshold for confident signal detection. 
Figure~\ref{fig:bayes} shows the detection boundaries obtained from 180 power-law SGWB injections, with combinations of spectral index $\alpha$ and amplitude $\Omega$ defined on a regular grid. The boundaries are interpolated between the simulated grid points.

For each injection, we run both the noise-only and noise-plus-signal models and estimate the corresponding evidence using stepping-stone sampling with $30$ temperatures. We chose this number after finding that increasing it from $30$ to $35$ produced nearly unchanged evidence estimates. For each model, we run five independent Markov chains with $2500$ iterations each, using the first $2000$ iterations for warm-up and retaining the remaining $500$ samples per chain, for a total of $2500$ posterior samples. For all four configurations, the uncertainty in the $\log_{10}B$ estimates is below $0.19$ for SNR values below 100.

Figure 2 compares the detection boundaries for the four combinations of SGWB model and test-mass noise prior considered in this work. For all four configurations, signals with sufficiently low amplitudes remain below the detection threshold. As $\Omega$ increases, the SGWB contributes more power to the data and becomes easier to distinguish from the instrumental noise, causing the Bayes factor to increase until it exceeds $B_\text{thresh}$.

The detection boundary depends on both $\Omega$ and $\alpha$. For negative spectral indices, more of the SGWB power is concentrated at low frequencies, where the LISA noise is dominated by test-mass acceleration noise. As shown in Appendix~\ref{appendix: sensitivity}, the $T$ channel is weakly sensitive to TM noise and therefore provides little information for constraining the TM noise spline. As a result, a flexible TM spline can absorb part of the weak SGWB signal, making it harder to separate from the noise.

This effect is most apparent for the agnostic signal model with the wide TM prior, \Magnoswide. In this case, both the signal and TM-noise models have considerable freedom, and higher SNRs are generally required for detection than when the informed TM prior is used. Interestingly, because the injected signals are power laws, the parametric-wide model \Mparamwide can detect signals at lower SNRs than the agnostic-wide model \Magnoswide, despite using the same noise priors. This difference is expected: the spline model has more parameters than the power-law model and therefore incurs a larger Occam penalty in the Bayes factor. Since its additional flexibility provides little improvement in fit for power-law injections, the evidence gap mainly reflects this penalty rather than a difference in fit.

Using the informed TM prior limits the freedom of the TM spline, making it less likely to absorb the SGWB signal. This increases the detectable region for both the parametric and agnostic signal models. Among
the four pipelines, the parametric model with the informed TM prior,
\Mparaminfo, gives the largest detectable region in the $(\alpha,\Omega)$ parameter space.

For positive spectral indices, more of the SGWB power lies at higher frequencies, where OMS noise is dominant. The $T$ channel provides useful information to constrain the OMS noise spline, thereby reducing the degeneracy between the signal and instrumental noise. As a result, the detection boundary generally lies closer to lower-SNR contours than it does in the negative-spectral-index region.

The detection boundary does not exactly follow a single SNR contour. This shows that SNR alone does not determine whether a signal is detected. The Bayes factor also depends on the spectral shape of the signal, the
flexibility of the signal model, and the prior information used for the instrumental noise.
\subsection{Gaussian Bump Analysis}
\label{gaussian_bump}
\begin{figure*}
    \centering
    \includegraphics[width=\textwidth]{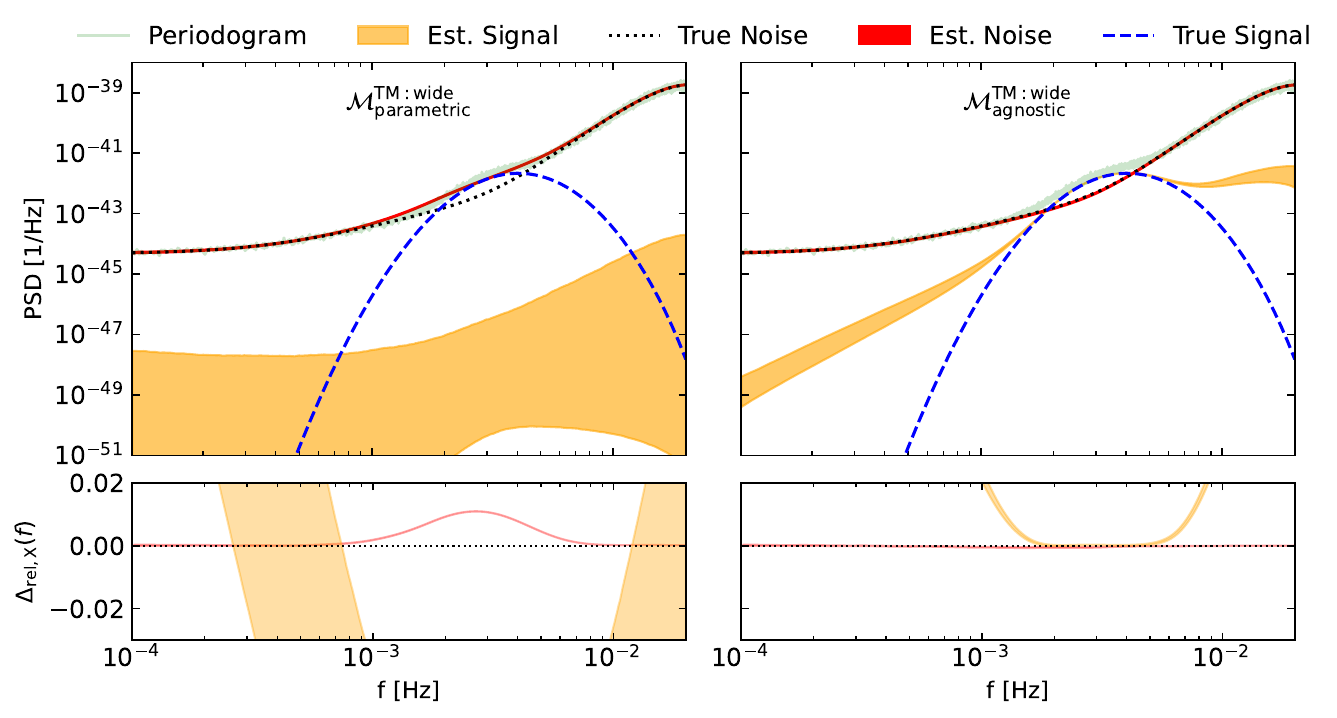}
    \caption{ Posterior spectral-density estimates for the Gaussian-bump injection obtained with the parametric-wide model, \Mparamwide (left), and the agnostic-wide model, \Magnoswide (right). In the upper panels, the gray curve shows the averaged periodogram, the dashed black and blue curves show the true instrumental-noise and injected SGWB spectra, and the red and orange shaded regions denote the corresponding $90\%$ posterior credible intervals. The lower panels show the relative deviations of the posterior quantiles from the true spectra, with the dotted horizontal line marking zero error.}
    \label{fig:gaussian_bump_mcmc}
\end{figure*}
In the study of cosmic backgrounds, a localized spectral feature, such as a Gaussian bump (GB), often represents evidence for specific primordial or early-universe phenomena. In contrast to astrophysical backgrounds that are often approximated by smooth power laws, such features can arise from phenomena including primordial black-hole formation and cosmological phase transitions~\cite{capriniReconstructingSpectralShape2019,flaugerImprovedReconstructionStochastic2021,nambaScaledependentGravitationalWaves2016}. A localized injection therefore provides a useful test of how well an inference framework can recover a signal whose spectral shape is not included in the assumed parametric model. It also allows us to assess the bias that may arise when an inappropriate signal model is imposed.

To test how the parametric and agnostic approaches perform when the signal shape is not known in advance, we inject a GB
\begin{equation}
    \Omega_{\text{gw}}(f) = \Omega_{*} \exp \left[ -\left( \frac{\log_{10}(f/f_{*})}{\sigma} \right)^2 \right],
\end{equation}
where $\Omega_{*}$ is the peak amplitude, $f_{*}$ is the central frequency, and $\sigma$ controls the width of the feature. We choose $\Omega_{*}=10^{-11}$, $\sigma=0.2$, and $f_{*}=3\times10^{-3}\,\mathrm{Hz}$, which places the feature near the most sensitive part of the LISA frequency band with SNR of $853.57$~\cite{capriniReconstructingSpectralShape2019}.

We convert this energy density into a strain PSD, $S_h(f)$, using the relation
\begin{equation}
    S_h(f) = \frac{3 H_0^2}{4 \pi^2 f^3} \Omega_{\text{gw}}(f).
\end{equation}
This strain PSD is then propagated through the LISA response function to yield the true signal spectrum.

Figure~\ref{fig:gaussian_bump_mcmc} shows the results for the GB injection using the power-law model \Mparamwide (left) and the fully spline model \Magnoswide (right). Because a power law cannot reproduce the bump, the parametric model fails to capture the feature. Instead, the signal estimate simply drops back down to the prior, forcing the flexible instrumental noise spline to absorb the spectral feature, making the noise estimates inaccurate as well.

In contrast, the agnostic model successfully separates the two components. It recovers the instrumental noise accurately and tracks the shape of the injected bump near its peak, where the signal is loudest. Further away from the peak, where the signal falls below the noise floor, the spline estimate becomes less accurate. Still, this example clearly shows the main advantage of the agnostic SGWB model, it can provide accurate estimates of spectral features that are not captured by an assumed parametric form, effectively reducing the bias caused by signal-model misspecification.
\section{Discussion}\label{conclusion}
In this work, we developed a Bayesian method to jointly estimate the LISA instrumental noise and the SGWB. We model the test-mass and optical metrology noise using log-P-splines and pass them through the TDI transfer functions. For the SGWB, we use either a power-law model or a log-P-spline model. A key advantage of the framework is that it uses a fixed-dimensional spline basis together with a roughness penalty. Since the number and positions of the knots are fixed before inference, the method avoids the need for trans-dimensional sampling methods such as reversible-jump MCMC. This keeps the analysis relatively simple and computationally efficient. We use this setup to study how the choice of signal model and the prior on the test-mass noise affect signal recovery and detection.

For the power-law injection with $\Omega=7\times10^{-14}$, $\alpha=2/3$, and ${\rm SNR}=18.3$, the power-law model gives tighter estimates. This is because the signal is well constrained in the most sensitive part of the band, roughly between $10^{-3}$ and $10^{-2}\,{\rm Hz}$, and the assumed power-law shape carries this information across the full frequency range. The spline signal model achieves comparable accuracy in the most sensitive part of the band, but it is more affected by the prior when the signal is weak.

Our analysis also provides a practical way to reduce the degeneracy between low-frequency SGWB signals and test-mass noise. Because the $T$ channel is only weakly sensitive to test-mass acceleration noise at low frequencies, it provides limited information for separating these two components. Our results show that an informative test-mass prior reduces this degeneracy by limiting how far the noise spline can move from its theoretical expectation. This increases the detectable region in the $(\alpha,\Omega)$ parameter space, with the largest improvement occurring for low-frequency-dominated signals. The Bayes-factor results further show that detectability is not determined by SNR alone. Signals with negative spectral indices place more power at low frequencies, where the degeneracy with test-mass noise is strongest. These signals generally require a higher SNR for detection when a wide TM prior is used. Signals with positive spectral indices place more power at higher frequencies, where the OMS noise is better constrained by the T channel. The detection boundary, therefore, depends on the signal shape, the flexibility of the signal model, and the prior used for the instrumental noise. These conclusions agree qualitatively with those of~\cite{karnesis_effect_2026}, who find with an independent pipeline that the prior adopted for the instrumental noise affects SGWB detectability more strongly than the choice between a parametric and a flexible signal model. Similarly, we find that the informed test-mass prior enlarges the detectable region for both signal models, while changing the signal model at a fixed noise prior has a smaller effect.
Since the two pipelines differ in spline basis, sampler, likelihood, and TDI channel treatment, this agreement is unlikely to be specific to one implementation.

The Gaussian-bump analysis shows the practical value of this flexibility. A power-law model cannot reproduce the localized feature, and part of the unmodeled signal is instead absorbed by the instrumental-noise spline. This affects both the signal and noise estimates. The flexible SGWB model performs much better in this case. It accurately recovers the instrumental noise spectrum and follows the injected bump around its peak, where the signal is most visible in the data. Although the signal remains weakly constrained away from the peak, this example shows that the spline model can recover spectral features that are not included in a standard power-law template.

Taken together, these results reveal a single trade-off. When the assumed signal shape is correct, the extra freedom of the spline model does not improve the fit but still costs prior volume, so the parametric model with the informed test-mass prior, \Mparaminfo, gives the largest detectable region in Fig.~\ref{fig:bayes}. When the assumed shape is wrong, that freedom prevents the signal from being absorbed into the instrumental-noise model, as shown by the Gaussian-bump injection in Fig.~\ref{fig:gaussian_bump_mcmc}. The choice of signal model is therefore a trade between sensitivity to anticipated signals and robustness to unanticipated ones.

There are several limitations to these results. We use an idealized LISA configuration, assuming equal arm lengths, stationary noise, and perfectly known response and transfer functions. However, the efficiency and flexibility of this joint spline framework make it a strong foundation for future work. Extending the model to handle realistic complexities, such as time-varying arm lengths, data gaps, or non-stationary noise, will likely require a time-frequency treatment, which we leave for future investigations. We also leave the case where the instrumental noise deviates from its theoretical expectation to future work.
\subsection*{Data and Software Availability}
The code developed for this work is publicly available at \url{https://github.com/nz-gravity/BOSS.git}. The repository contains the source code and example scripts required to reproduce the results presented in this work. The datasets generated for this study are available at Zenodo repository~\cite{aimenBayesianPsplineRecovery2026}. The LISA response functions used in this analysis were computed using the \texttt{backgrounds} software and the accompanying isotropic SGWB response implementation developed by Baghi et al.~\cite{baghi_backgrounds_response}. We performed the numerical calculations using \texttt{NumPy}~\cite{harrisArrayProgrammingNumPy2020} and \texttt{SciPy}~\cite{virtanenSciPy10Fundamental2020}. We generated the figures using \texttt{Matplotlib}~\cite{hunterMatplotlib2DGraphics2007}.
\begin{acknowledgments}
We are especially grateful to Quentin Baghi and Nikolaos Karnesis for their extensive help, valuable discussions, and feedback throughout this project. We additionally thank Quentin Baghi for providing the LISA response-function code used in this work. We also thank the members of the LISA Noise Non-Stationarities Group, part of the ``Deep Analysis Group'' of the Distributed Data Processing Centre (DDPC), for helpful discussions.

NA, NC, PMR, RM, and AV gratefully acknowledge support from the Marsden Fund Council grant MFP-UOA2531, funded by the New Zealand Government and managed by the Royal Society Te Apārangi. The work of NC, PMR, and  RM was further supported by the CNRS International Research Project (IRP) OG-Science FR-NZ, and NC received financial support from the CNRS MITI interdisciplinary programs and from the French Agence Nationale de la Recherche.

This work was performed on the OzSTAR national facility at Swinburne University of Technology. The OzSTAR program receives funding in part from the Astronomy National Collaborative Research Infrastructure Strategy (NCRIS) allocation provided by the Australian Government and from the Victorian Higher Education State Investment Fund (VHESIF) provided by the Victorian Government.
\end{acknowledgments}
\appendix
\section{Sensitivity of the TDI channels to instrumental noise perturbations}
\label{appendix: sensitivity}
\begin{figure*}[!t]
    \centering
    \includegraphics[width=\textwidth]{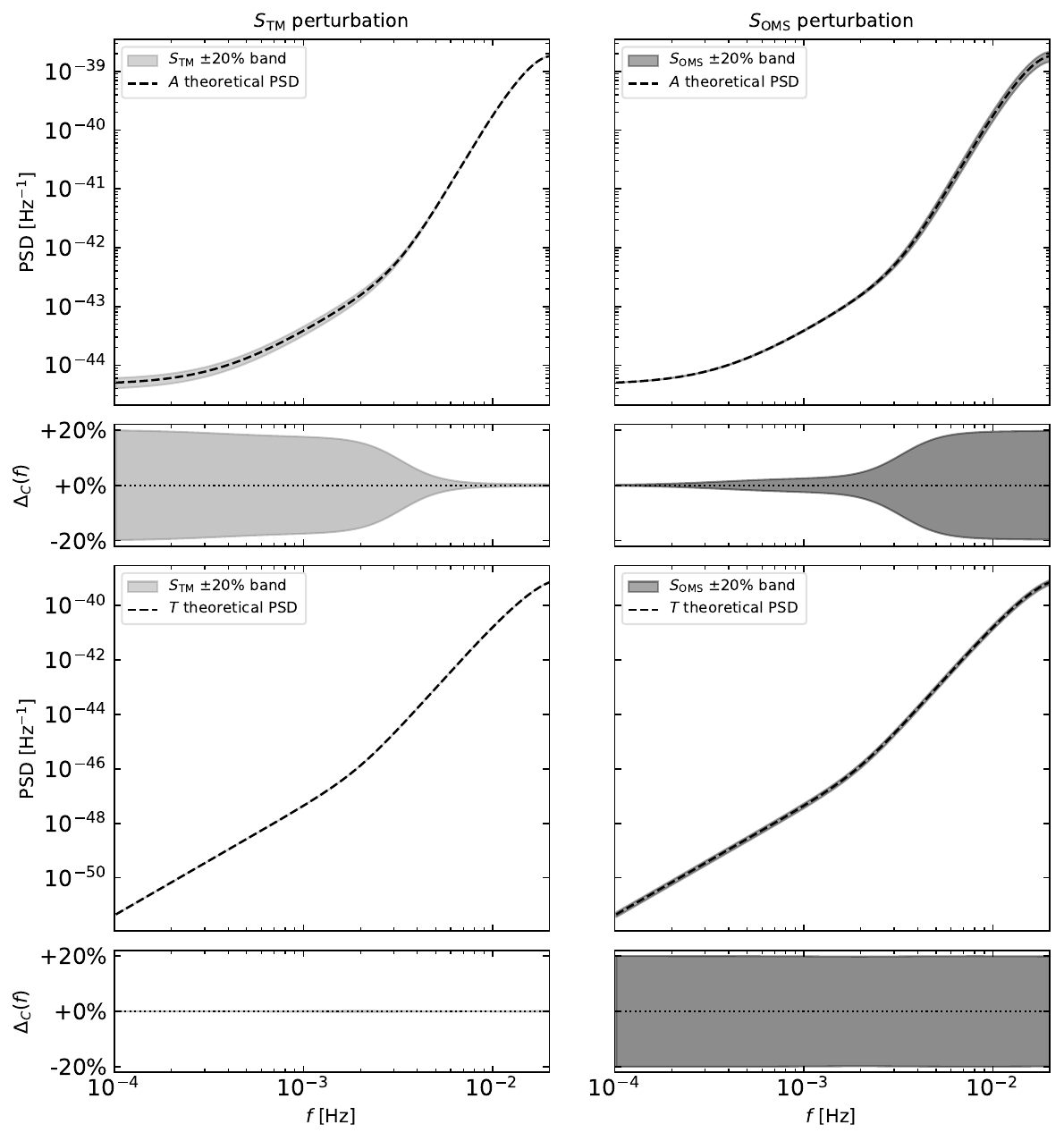}
    \caption{
    Effect of $\pm20\%$ perturbations in the test-mass noise and optical metrology noise on the $A$ and $T$ channel spectra. The black curves show the unperturbed spectra, and the shaded regions show the range obtained after perturbing one noise component at a time. The residual panels show the corresponding fractional change relative to the unperturbed spectrum.}
    \label{fig:A_pert}
\end{figure*}

In this appendix, we examine how perturbations in the test-mass and optical metrology noise levels affect the $A$ and $T$ channels. We do this to explore how the two noise components contribute at all frequencies, and how their relative importance changes after they are propagated through the TDI transfer functions. We use the same settings as in the main SGWB analysis.

We first perturb the test-mass and optical metrology noise spectra separately. The perturbations are defined as
\begin{align}
    S_{\rm TM}^{(\pm)}(f)  &= (1 \pm a)\,S_{\rm TM}(f), \\
    S_{\rm OMS}^{(\pm)}(f) &= (1 \pm a)\,S_{\rm OMS}(f),
\end{align}
where $a=0.2$, corresponding to a $\pm20\%$ change in one noise component at a time. For each channel $C\in\{A,T\}$, these perturbed noise spectra are then propagated through the analytic TDI transfer functions. This gives
\begin{align}
    S^{(\pm)}_{C,{\rm TM}}(f) &=
        S_{C}\bigl(f; S_{\rm TM}^{(\pm)}(f), S_{\rm OMS}(f)\bigr), \\
    S^{(\pm)}_{C,{\rm OMS}}(f) &=
        S_{C}\bigl(f; S_{\rm TM}(f), S_{\rm OMS}^{(\pm)}(f)\bigr).
\end{align}
Here, $S_{C}\bigl(f; S_{\rm TM}(f), S_{\rm OMS}^{(\pm)}(f)\bigr)$ denotes the channel spectrum obtained from the TM and OMS noise components. The first expression varies only the test-mass noise, while keeping the optical metrology noise fixed. The second expression varies only the optical metrology noise, while keeping the test-mass noise fixed.

Figure~\ref{fig:A_pert} shows the resulting perturbation bands for the $A$ and $T$ channel spectra. The black dashed curves show the unperturbed spectra, while the shaded regions show the range obtained from the $+20\%$ and $-20\%$ perturbations. To quantify these changes, we also compute the relative residual
\begin{equation}
    \Delta_{C}(f)
    = \frac{S_{C}^{(\pm)}(f) - S_{C}(f)}{S_{C}(f)}.
\end{equation}
These residuals are shown below the corresponding spectra in Fig.~\ref{fig:A_pert}. For the $A$ channel, perturbing the test-mass noise mainly affects the low-frequency part of the spectrum, while perturbing the optical metrology noise mainly affects the high-frequency part. For the $T$ channel, the behavior is different: perturbing the test-mass noise produces a negligible change, while perturbing the optical metrology noise changes the spectrum across the full frequency range. This shows that, for this transfer-function model and frequency range, the $T$ channel is mainly sensitive to optical metrology noise.

\begin{figure}[!t]
    \centering
    \includegraphics[width=\columnwidth]{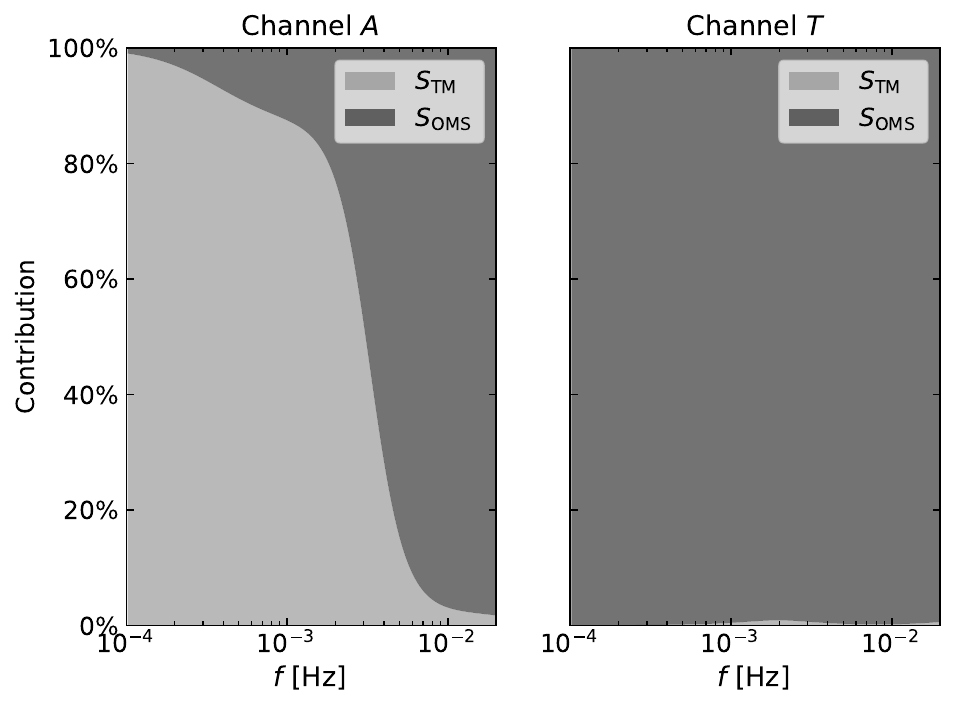}
    \caption{
    Fractional contributions of test-mass noise and optical metrology noise to the $A$ and $T$ channel spectra. The $A$ channel is test-mass-dominated at low frequencies and optical-metrology-dominated at high frequencies. The $T$ channel is almost entirely dominated by optical metrology noise over the frequency range used in this analysis.}
    \label{fig:AT_contri}
\end{figure}

We next look at the fractional contribution of each instrumental noise component to the total TDI spectrum. For channel $C$, we write
\begin{align}
    p^{\rm TM}_C(f)
      &= \frac{S^{\rm TM}_{C}(f)}{S_{C}(f)}, \\
    p^{\rm OMS}_C(f)
      &= \frac{S^{\rm OMS}_{C}(f)}{S_{C}(f)}
       = 1 - p^{\rm TM}_C(f),
\end{align}
where $S^{\rm TM}_{C}(f)$ and $S^{\rm OMS}_{C}(f)$ are the test-mass and optical metrology parts of the channel spectrum.

Figure~\ref{fig:AT_contri} shows these fractional contributions for the $A$ and $T$ channels. In the $A$ channel, the test-mass noise dominates at low frequencies, while the optical metrology noise dominates at high frequencies. The transition between the two regimes occurs in the middle of the frequency range considered here. In contrast, the $T$ channel is almost entirely dominated by optical metrology noise across the full band, with only a very small test-mass contribution. This is consistent with the perturbation study in Fig.~\ref{fig:A_pert}, where changing the test-mass noise has almost no effect on the $T$ channel spectrum.

These results are important for the noise and SGWB inference. The $T$ channel can help constrain the optical metrology noise because it is highly sensitive to it. However, it provides little information about the test-mass noise in the low-frequency range. The test-mass noise must therefore be constrained mainly either through the $A$ channel or through a prior. This affects the analysis of low-frequency-dominated SGWB. This is particularly important for low-frequency-dominated SGWB signals, where a flexible noise model can lead to identifiability issues. In such cases, a tighter prior on the test-mass noise can help improve SGWB recovery.
\section{Prior width}\label{appendix: phi}
\begin{figure}[!t]
    \centering
    \includegraphics[width=\columnwidth]{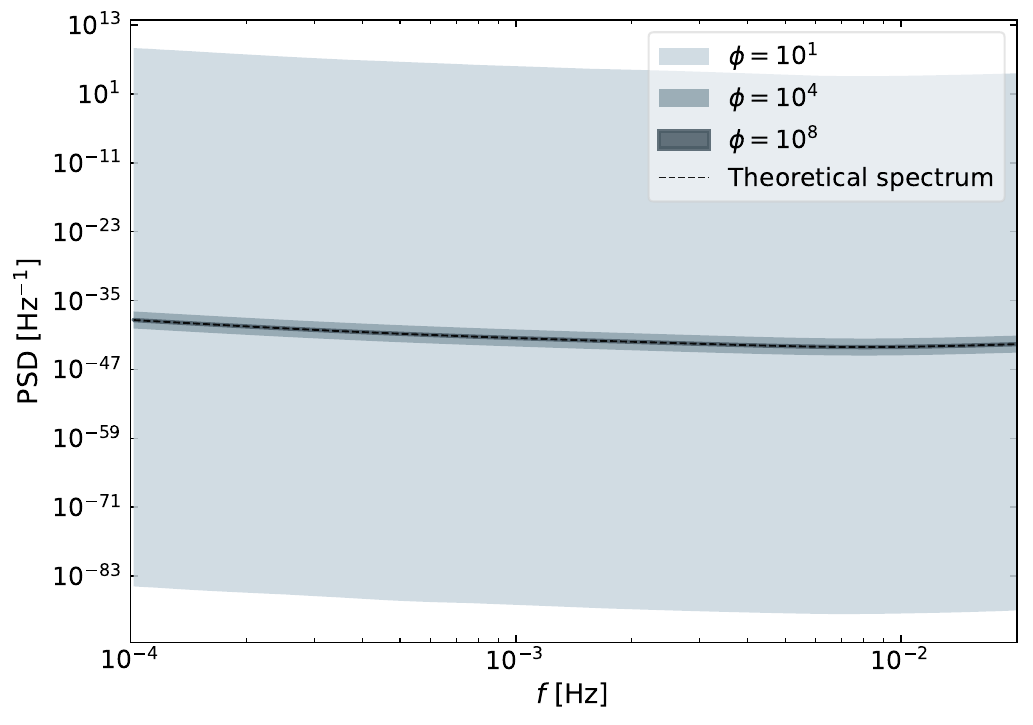}
    \caption{Effect of different precision parameter values, $\phi$, on the 90\% prior credible region for the PSD. The dashed black curve represents the theoretical spectrum used as the prior's center.}
    \label{fig:priorwidthphi}
\end{figure}
This appendix illustrates the behavior of the log-P-spline prior distribution. Figure~\ref{fig:priorwidthphi} shows the pointwise quantiles of the prior PSD calculated from random samples, with the dashed black line representing the baseline theoretical spectrum used as the center of the prior. The precision parameter $\phi$ controls how tightly the spline weights are concentrated around this theoretical spectrum. A smaller value of $\phi$ results in a wider prior, giving the PSD more freedom to deviate from the theoretical curve. Conversely, a larger value of $\phi$ produces a narrower prior, concentrating the sample paths more tightly around the baseline spectrum. The shaded region in Fig.~\ref{fig:priorwidthphi} represents the 5th--95th percentile range and therefore does not show the full range of spectra allowed by the prior. Because the spline coefficients have Gaussian priors, the distribution is not bounded by these quantiles, and individual realizations can extend beyond the displayed region. This is particularly relevant for the wide prior, where greater freedom in the noise spectrum can increase the degeneracy between instrumental noise and the SGWB signal. This behavior motivates the choice of prior widths used in our main analysis. The wide prior ($\phi=10^4$) allows greater flexibility in the noise model but can absorb part of the SGWB signal, whereas the informed prior ($\phi=10^8$) restricts the noise model more strongly and helps reduce this degeneracy.
\bibliographystyle{unsrt}
\bibliography{biblio}
\end{document}